\documentstyle[graphicx,psfig]{mn}

\newcommand{\etal}{et al.\ }
\newcommand{\ie}{i.e.\ }
\newcommand{\eg}{e.g.,\ }
\newcommand{\GC}{Galactic Centre}
\newcommand{\GCR}{Galactic Centre Region}

\newcommand{\NH}{$N_{\rm H}$}
\newcommand{\NHUNIT}{H~cm$^{-2}$}
\newcommand{\FX}{$F_{\rm X}$}
\newcommand{\FLUXUNIT}{erg~s$^{-1}$~cm$^{-2}$}
\newcommand{\LX}{$L_{\rm X}$}
\newcommand{\LUMIUNIT}{erg~s$^{-1}$}

\newcommand{\ASCA}{\it ASCA}
\newcommand{\SAX}{\it SAX}

\newcommand{\RXTE}{\it RXTE}
\newcommand{\XMM}{\it XMM}
\newcommand{\XMMN}{\it XMM-Newton}
\newcommand{\Chandra}{\it Chandra}

\newcommand{\XMMfe}{XMM J174544$-$2913.0}
\newcommand{\XMMtr}{XMM J174457$-$2850.3}

\newcommand{\hms}[3]{$#1^{\rm h}~#2^{\rm m}~#3^{\rm s}$}
\newcommand{\hmss}[4]{$#1^{\rm h}~#2^{\rm m}~#3\fs#4$}

\newcommand{\dms}[3]{$#1\degr~#2\arcmin~#3\arcsec$}
\newcommand{\dmss}[4]{$#1\degr~#2\arcmin~#3\farcs#4$}

\title{Unusual X-ray transients in the Galactic Centre}
\author[M. Sakano, R.S. Warwick, A. Decourchelle and Q.D. Wang]
       {M. Sakano$^1$, R. S. Warwick$^1$, A. Decourchelle$^2$ 
and Q. D. Wang$^3$\\
        $^1$Department of Physics and Astronomy, University of Leicester,
	Leicester LE1 7RH, UK\\
	$^2$CEA/DSM/DAPNIA, Service d'Astrophysique, C.E. Saclay, 
	91191 Gif-sur-Yvette Cedex, France\\
	$^3$Astronomy Department, University of Massachusetts, Amherst, USA 
}

\date{Accepted XXXX. Received XXXX 2003}

\pagerange{\pageref{firstpage}--\pageref{lastpage}}
\pubyear{2003}

\begin{document}

\maketitle

\label{firstpage}

\begin{abstract}

We report the discovery  in the {\GC} region of two hard X-ray sources, 
designated as {\XMMtr} and {\XMMfe}, which exhibited flux variations in the 
2--10 keV band in excess of a factor of 100  in observations spanning roughly 
a year. In both 
cases the observed hydrogen column density is consistent with a 
location near to the  {\GC}, implying peak X-ray luminosities of 
$\sim 5 \times 10^{34}$ {\LUMIUNIT}.  These objects may represent a new 
population of transient source 
with very different properties to the much more luminous {\GC} 
transients associated with neutron star and black-hole binary systems. 
Spectral analysis shows that
{\XMMtr} has relatively weak iron-line emission set against a very hard 
continuum. {\XMMfe}, on the other hand, has an extremely strong K-line 
from helium-like iron with an equivalent width of $\sim$2.4~keV.
The nature of the latter source is of particular interest. Does it represent
an entirely new class of object or does it correspond to a known class 
of source in a very extreme configuration?
\end{abstract}

\begin{keywords}
Galaxy:centre -- X-rays:binaries -- X-rays:individual:XMM J174457$-$2850.3 -- X-rays:individual:XMM J174544$-$2913.0
\end{keywords}

\section{Introduction}

During the last four years, the {\GC} region has been observed repeatedly by 
{\Chandra} and {\XMMN}, including several extensive campaigns focused 
on Sgr~A$^*$ ({\eg} Baganoff {\etal} 2001, 2003; 
Goldwurm {\etal} 2003; Porquet {\etal} 2003). 
Both observatories have also carried out wide-angle 
X-ray surveys in which sets of overlapping pointings 
have been used to give coverage of the Galactic plane within $1^{\circ}$ 
of the {\GC} (Wang, Gotthelf \& Lang 2002; Warwick 2002; Sakano {\etal} 2004c).
% Decourchelle {\etal} 2004).
The excellent imaging capability and high sensitivity of the two
observatories has allowed the discrete X-ray source population in the
direction of the {\GC} to be studied over a range in X-ray luminosity 
extending from $10^{38}$ {\LUMIUNIT} down to $\sim 10^{31}$ {\LUMIUNIT}.
The temporal and spectral properties of {\GC} low-mass and high-mass
X-ray binaries, typically seen in outburst with $L_{\rm X} > 10^{36}$ 
{\LUMIUNIT},  are now well established through the work of 
numerous missions, past and present, including most recently {\RXTE}.
However, much less is known about {\GC} sources with peak luminosities below 
$\sim 10^{35}$ {\LUMIUNIT}, since this is close to the effective 
detection limit for earlier hard X-ray imaging missions such as 
{\ASCA} and {\SAX}.  In essence, the recent surveys of
{\Chandra} and {\XMMN} have provided a new window on faint source populations
at the {\GC} ({\eg} Muno {\etal} 2003a, 2003b, 2004b).

The spectra of the X-ray sources in the {\GCR} with $L_{\rm X} > 10^{35}$ {\LUMIUNIT}, 
most of which are low-mass X-ray binaries (LMXB)
containing either a neutron star or black hole, are often featureless 
(Sidoli {\etal} 1999; Sakano {\etal} 2002).  On the other hand, Wang {\etal} 
(2002) report that the summed spectrum of the faint sources detected in 
the {\Chandra} wide-angle {\GC} Survey shows significant 6.7 keV
Fe K-line emission, implying the existence of one or more different 
populations of sources at lower luminosity. More recently, the $\sim 2000$ 
X-ray sources detected in the very deep {\Chandra} observations of the field 
around Sgr~A$^*$, have been shown on average to have very hard (photon
index $\Gamma < 1$) 
spectra in addition to strong He-like and H-like K-lines from Si, S, Ar, 
Ca and Fe (Muno {\etal} 2003a, 2004b).  Although resolved point sources 
contribute up to $\sim 10\%$ of total hard X-ray emission from the {\GC},
the bulk of the X-ray luminosity of the region must either be truly 
diffuse in nature or originate in a very faint population
of point sources with very hard spectra which are more numerous than 
cataclysmic variables (Muno {\etal} 2003a, 2004a). In the case of
the former, it remains unclear whether thermal or non-thermal 
process dominate and what is the energising source of the plasma.

Comparing {\GC} observations taken at different epochs it is straightforward
to pick out highly variable or transient X-ray sources. In this paper
we present the results for two such transient sources with peak observed
luminosities of $\sim 5 \times 10^{34}$ {\LUMIUNIT}, which have 
interesting spectral properties. We also comment on what contribution such
sources, in relative quiescence, might make to the unresolved 
X-ray hard emission of the {\GC}.

\section[]{Observations}

We have searched for transient X-ray sources by comparing images 
from  the wide-angle X-ray surveys of the {\GC} region carried out by 
{\Chandra} and {\XMMN} (Wang {\etal} 2002; Warwick 2002; Sakano, Warwick 
\& Decourchelle 2003).
%Decourchelle {\etal} 2004).
The results of a full analysis will be discussed in a later paper
and here we confine our attention to two relatively bright transients with 
interesting spectral properties.
 
\begin{figure*}
\centering
\begin{minipage}{0.75\textwidth}
\centering\hbox{\includegraphics[width=12 cm]{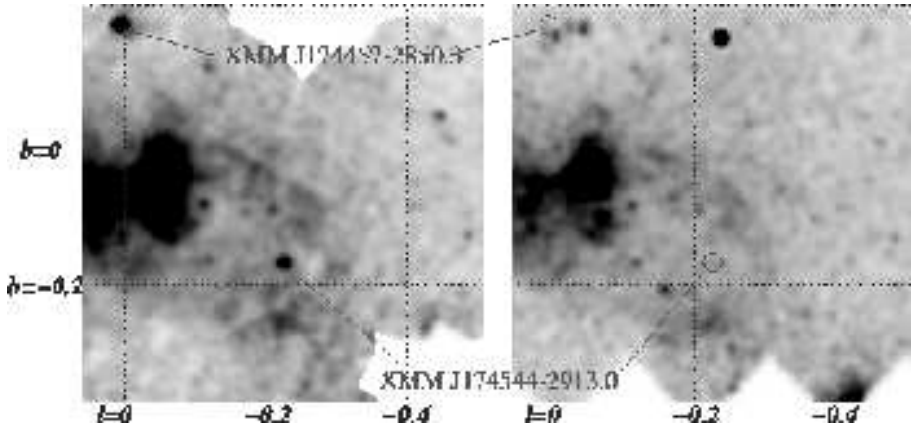}}
\caption{Sub-regions from the {\XMMN}/MOS1+2 ({\it left panel}) and 
{\Chandra}/ACIS  ({\it right panel}) mosaiced images of the {\GCR}.
The respective energy bands are 2--9 keV and 1--8 keV.
Both images are background subtracted as well as vignetting and exposure 
corrected and have been smoothed with a circular Gaussian mask 
with $\sigma=20''$. The grey-scaling is set such that regions of extended 
emission appear the same in the two images. The two transient sources 
which are the focus of the current paper are indicated.  The bright 
extended feature on the eastern edge of the field of view is the 
Sgr~A region ({\eg} Maeda {\etal} 2002; Sakano {\etal} 2004b).
}
\label{fig:xmm_cxo_img}
\end{minipage}~~
\end{figure*}

Fig.~\ref{fig:xmm_cxo_img} shows a sub-region from the image mosaics
produced from the {\XMMN} and {\Chandra} {\GC} surveys.  The locations of 
the two transient sources, designated as {\XMMtr} and {\XMMfe}, are 
indicated.  Both sources are relatively bright in the {\XMMN} map
but not visible in the corresponding {\Chandra} image.
Based on the {\XMMN} data, their positions (RA, DEC, J2000) are determined 
to be
(\hmss{17}{44}{57}{34}, \dmss{-28}{50}{20}{3})  and
(\hmss{17}{45}{44}{51}, \dmss{-29}{13}{0}{6}),
respectively, with an error radius of 4 arcsec.

Having identified the two sources of interest, we focus on the individual 
{\XMMN} and {\Chandra} observations which encompass their positions
(see  Table~\ref{tbl:obslog}).
In the case of {\XMMN}, the data are from the EPIC instrument, which consists
of three cameras, two utilising MOS CCDs (Turner {\etal} 2001) and one 
employing a pn CCD (Str\"{u}der \etal 2001).  In the present observations
the EPIC MOS and pn cameras were operated in {\it Full Frame Mode} and 
{\it Extended Full Frame Mode}, respectively, with the medium filter selected.
We have used the Standard Analysis Software ({\sc SAS}) Version 5.3 for the 
data filtering and reduction.  Observing periods affected by high levels
of background flaring identified in the light curve of the
full-field data above 10~keV were rejected. For the MOS cameras we utilised 
pixel patterns of 0--12 (single to quadruple), whereas for the pn camera, 
we accepted single and double (pattern 0-4) events.  All the {\Chandra}/ACIS 
observations were carried out in FAINT mode, with the optical axis 
located at the nominal position of ACIS-I. The data reduction and filtering 
were performed with the Chandra Interactive Analysis Software ({\sc ciao}) 
Version 2.3.  We used events with the single to quadruple pixel patterns.
The {\Chandra} observations targeted on Sgr~A$^*$ (Table~\ref{tbl:obslog}), 
carried out from 22 May to 4 June, 2002 have essentially the same attitude
and for the spectral and image analysis considered in \S\ref{sec:XMMtr} were treated 
as one dataset.

\begin{table*}
 \centering
 \begin{minipage}{140mm}
  \caption[]{{\XMMN} and {\Chandra} observations used in the present 
analysis. \label{tbl:obslog}
}
\begin{tabular}{lccccc}
   \hline
 Name & Obs-ID & RA & DEC & Obs. Date & Net Exp.\footnote{The net exposure 
time after filtering - EPIC MOS1/MOS2/pn in the case of {\XMMN} and 
ACIS-I in the case of {\Chandra}.} 
\\
      &        & (J2000) & (J2000) & (yyyy/mm/dd) & (ks) \\
   \hline
   \hline
   \multicolumn{6}{c}{\XMMN}\\
   \hline
GC10 & 0112971001 & \hms{17}{46}{40} & \dms{-29}{13}{00} & 2000/09/24 & 9.5/9.5/7.8 \\
GRO J1744$-$28 & 0112971901 & \hms{17}{44}{34} & \dms{-28}{45}{22} & 2001/04/01 & 8.7/8.9/4.1\\
 GC6 & 0112972101 & \hms{17}{45}{40} & \dms{-29}{00}{23} & 2001/09/04 & 22.4/24.0/17.5\\ % (Rev-ID=318)
   \hline
   \hline
   \multicolumn{6}{c}{\Chandra}\\
   \hline
%SGR A$^*$ & &  242 & \hms{17}{45}{39} & \dms{-29}{00}{47} & 1999/09/21 & 38.7 \\
GCS 12    & 2279 & \hms{17}{45}{19} & \dms{-28}{40}{22} & 2001/07/18 & 11.6\\
GCS 14    & 2284 & \hms{17}{45}{37} & \dms{-28}{56}{27} & 2001/07/18 & 10.6\\
GCS 15    & 2287 & \hms{17}{44}{51} & \dms{-28}{50}{21} & 2001/07/18 & 10.6\\
GCS 16    & 2291 & \hms{17}{45}{55} & \dms{-29}{12}{34} & 2001/07/18 & 10.6\\
GCS 17    & 2293 & \hms{17}{45}{09} & \dms{-29}{06}{28} & 2001/07/19 & 11.1\\
GCS 20    & 2267 & \hms{17}{44}{41} & \dms{-29}{16}{27} & 2001/07/19 & 11.5\\
SGR A$^*$ & 2943 & \hms{17}{45}{41} & \dms{-29}{00}{15} & 2002/05/22 & 38.0\\
SGR A$^*$ & 3663 & \hms{17}{45}{41} & \dms{-29}{00}{15} & 2002/05/24 & 38.0\\
SGR A$^*$ & 3392 & \hms{17}{45}{41} & \dms{-29}{00}{15} & 2002/05/27 & 166.7\\
SGR A$^*$ & 3393 & \hms{17}{45}{41} & \dms{-29}{00}{15} & 2002/05/30 & 158.0\\
SGR A$^*$ & 3665 & \hms{17}{45}{41} & \dms{-29}{00}{15} & 2002/06/03 & 89.9\\
   \hline
\end{tabular}
\end{minipage}
\end{table*}

\section[]{Results}
 \label{sec:result}

\subsection[]{{\XMMtr}}
 \label{sec:XMMtr}

The location of {\XMMtr} lies within the field of view of 10 of the 
pointings listed in Table~\ref{tbl:obslog}. Table~\ref{tbl:flux} 
summarises, in chronological order, the various flux determinations at 
the source position. Here we quote 3$\sigma$ upper limits
when the source was not detected and 1$\sigma$ errors on
significant detections using the 2--8 keV band
image.

\begin{figure}
\psfig{file=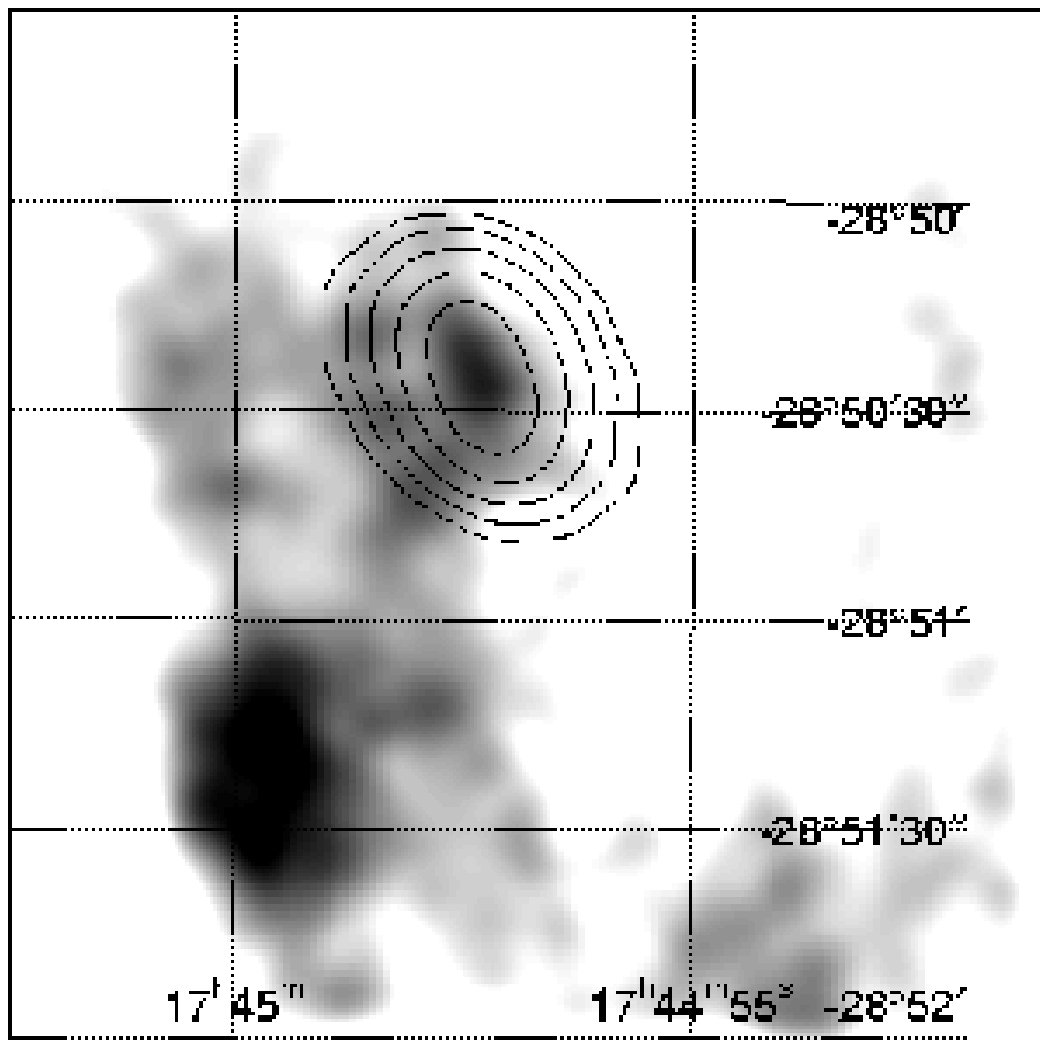,width=0.45\textwidth,clip=}
\caption{Chandra 2--8 keV image of the {\XMMtr} region based on the 
combined 2002 data. The contours represent the point spread function (PSF) 
at this off-axis location.   Both the image and the PSF are smoothed with a
Gaussian filter with $\sigma=3''$.  The grey-scale and contour 
levels are logarithmically spaced. 
 \label{fig:xmmtr-imgpsf}}
\end{figure}

This source is most clearly detected in the {\XMMN} 
GC6 observation carried out on 2001/9/04,
while two months earlier and ten months later the source was found to
be 40--400 times fainter (Table~\ref{tbl:flux}).
Fig.~\ref{fig:xmmtr-imgpsf} shows an image of the source region 
obtained from the combined 2002 {\Chandra} data when the source 
was in a particularly low state. The overlaid contours represent the 
point spread function (PSF) at the position of a local peak 
which, within the errors, is coincident with the {\XMMN}
position\footnote{The local peak is detected at (RA, Dec) = 
(\hmss{17}{44}{57}{3}, \dmss{-28}{50}{26}) within an error radius of 
$10''$;  the large uncertainty in the position is due to the 13.6~arcmin 
offset from the optical axis.}.
There are low-surface brightness features to the south and east of this
peak.
Comparing the
photon statistics of the peak and its surroundings, we estimate the
significance of the point source in these observations to 
be 7.4$\sigma$.

When the source was in its high state (GC6 observation:
Table~\ref{tbl:flux}), there was evidence for low-amplitude temporal
variations.  For example a trial fitting of the light curve, with a
constant emission model (after the background is subtracted), for a
binning of 32~s, is rejected at more than 99.9\% confidence via
$\chi^2$-test.  On the other hand no distinct burst, flare, or dip
behaviour was apparent.  Extracting a power density spectrum reveals a
slight excess of power above the Poisson noise level at $\sim$0.06~Hz
and at frequencies below $\la$0.005~Hz.  A search for pulsations using a
fast Fourier transformation (FFT) algorithm showed evidence of a peak at
$\sim$0.19~Hz.  An epoch folding search gave the best-fitting period to
be 5.2521~s with a nominal error ($P^2/t$) of 0.0012~s.  The folded
light curve with this period shows a sinusoidal form with a maximum
relative amplitude of 20\%.  However, given the relatively poor signal
to noise ratio, the reliability of this putative periodic signal 
is very uncertain.

%\subsubsection[]{Spectrum}
% \label{sec:XMMtr-sp}

\begin{figure}
%\begin{figure*}
\centering
%\begin{minipage}{0.67\textwidth}
\centering\hbox{\includegraphics[width=5.5 cm, angle=270]{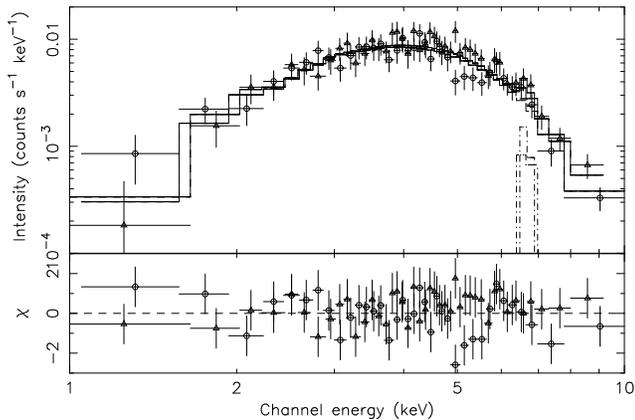}}
 %\vspace{150pt}
\caption{EPIC MOS spectra of {\XMMtr} from the September 2001 (GC6) 
observation. The open circles and triangles represent MOS1 and
MOS2 data, respectively.  The best-fitting model is 
shown as a histogram. The lower panel shows the residuals to the best-fitting
model.
}
\label{fig:xmmtr-sp-xmm}
%\end{minipage}~~
%\end{figure*}
\end{figure}

\begin{figure}
%\begin{figure*}
\centering
%\begin{minipage}{0.67\textwidth}
\psfig{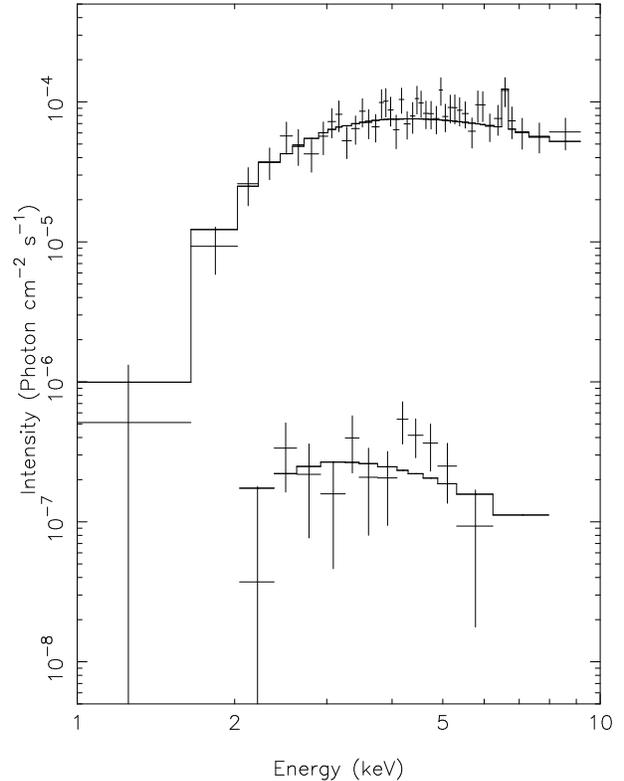}
 %\vspace{150pt}
\caption{Unfolded X-ray spectra of {\XMMtr}.  The upper and lower data are 
taken with {\XMMN} (MOS2) in 2001  and with {\Chandra} (ACIS-I) in 2002, 
respectively.
}
\label{fig:xmmtr-sp}
%\end{minipage}~~
%\end{figure*}
\end{figure}

The spectrum of the source as recorded by the {\XMMN} MOS cameras in the 
GC6 observation was extracted using an elliptical region of semi-axis
dimensions $52''$ and $18''$ centred on the source position
and aligned with the PSF ({\it nb.} this cell shape and alignment was in part
dictated by the position of the source at the edge of the MOS field of 
view). Corresponding background spectra were taken from a nearby sky region 
of the same shape and orientation as the source region, chosen
so as to avoid the weak sources detected in the corresponding
{\Chandra} observations. The source was outside of the field of view
of the EPIC pn camera.

Fig.~\ref{fig:xmmtr-sp-xmm} shows the {\XMM} MOS1 and MOS2 spectra after
background subtraction.  The observed spectrum is relatively featureless except
for a possible excess signal around 6--7 keV.  We fitted the two MOS spectra
simultaneously with a power-law component modified by interstellar
absorption, and found that this simple model well approximated the data 
($\chi^2$/dof=63.8/77).  Nevertheless, when a narrow
Gaussian line was added in the 6--7 keV range the fit
improved ($\chi^2$/dof=58.9/75) with a significance of 95\% as measured
by the F-test\footnote{The F-test may underestimate the significance in
this case (Protassov {\etal} 2002).  However the result here that the
line is significant would be unchanged.}.  The line equivalent width was
found to be 180 eV 
with a 90\% confidence interval  of 40--320 eV (90\% confidence
intervals assuming one interesting parameter are used throughout 
this paper).  The best-fitting model and parameters
including the line are shown in Fig.~\ref{fig:xmmtr-sp-xmm} and
Table~\ref{tbl:fit}, respectively.
On the basis
of the line energy measurement, the line is very likely to be the 
6.7-keV K-line from helium-like iron.  To check the contribution of the
detector and diffuse background to this line we tried a 50\% larger 
normalisation for the background data; the result was that the significance 
and the intensity of the line were hardly affected.
 The observed  2--10~keV flux of 4.5
$\times 10^{-12}${\FLUXUNIT} converts to an absorption-corrected luminosity of
4.5$\times 10^{34}${\LUMIUNIT}, assuming a distance of 8.0~kpc (Reid 1993). 

A remarkable feature of the source spectrum is the apparent hardness of the 
continuum.  In effect, this excludes a thermal bremsstrahlung model -- an 
actual fit gave a 90\% lower limit for the temperature of such a component to 
be 35~keV.

We have also analysed the ACIS-S spectrum  of {\XMMtr} as obtained from
the co-added {\Chandra} data of 2002 May/June.  
  In order to obtain the maximum 
signal-to-noise ratio and to minimize the uncertainty due to possible
spatial variation of the
diffuse component (see Fig.~\ref{fig:xmmtr-imgpsf}),
we chose a relatively small region for the
accumulation of the source spectrum, namely an elliptical region
encompassing the central part of the PSF, with semi-axes of $10''$ and
$5''$.  We then collected the background spectrum from a region east
of the source position, where some diffuse emission extending from the
source region is seen.

The background-subtracted {\Chandra} spectrum was modelled with a 
power-law continuum with the low-energy absorption fixed at the value obtained
from the {\XMMN} high-state measurement.  The derived best-fitting photon 
index was 1.9 (1.1--2.8) (Table~\ref{tbl:fit}). In this case the addition of an iron line at 
6.7 keV did not improve the fit with a 90\% upper limit on the equivalent 
width of such a line determined to be 760~eV.  The observed 2--10 keV 
flux is 1.0$\times 10^{-14}${\FLUXUNIT}, corresponding to an absorption-corrected luminosity 
of 1.2$\times 10^{32}${\LUMIUNIT}. Fig.~\ref{fig:xmmtr-sp} compares the 
unfolded spectra of {\XMMtr} in the high and low states.  

\begin{table*}
 \centering
 \begin{minipage}{140mm}
  \caption[]{The measured X-ray flux and derived X-ray luminosity of the 
two sources. \label{tbl:flux}
}
  \begin{tabular}{lclcccc}
   \hline
   Date & Satellite & Field & \multicolumn{2}{c}{\XMMtr} & \multicolumn{2}{c}{\XMMfe}\\
    &  &  & {\FX}\footnote{The observed 2--10 keV flux in units of
   10$^{-13}${\FLUXUNIT}.} & {\LX}\footnote{The 2--10 keV
   absorption-corrected luminosity in units of 10$^{34}$ {\LUMIUNIT}.} & 
{\FX}$^{a}$ & {\LX}$^{b}$ \\
   \hline
   2000/09/24 & {\XMM}     & GC10       & --- & --- & 37 & 5 \\
   2001/04/01 & {\XMM} & GRO J1744$-$28 & $<$1.6 & $<$0.2 & --- & --- \\
   2001/07/18 & {\Chandra} & GCS 12     & 1.2$\pm$0.2 & 0.12$\pm$0.02 & --- & --- \\
   2001/07/18 & {\Chandra} & GCS 14     & $<$0.5 & $<$0.05 & --- &  --- \\
   2001/07/18 & {\Chandra} & GCS 15     & $<$3.1 & $<$0.3 & --- &  --- \\
   2001/07/18 & {\Chandra} & GCS 16     & --- & --- & $<$0.3\footnote{The position of
   {\XMMfe} is close to a chip gap ($\sim$15 arcsec).  If the true
   source position is located closer to the chip gap than our
   current estimate, these upper limits may have to be increased by up to 50\%.}                     & $<$0.04$^{c}$ \\ 
   2001/07/19 & {\Chandra} & GCS 17 & --- & --- & $<$0.4 & $<$0.05  \\
   2001/07/19 & {\Chandra} & GCS 20 & --- & --- & $<$0.8 & $<$0.11 \\
   2001/09/04 & {\XMM} & GC6 & 45 & 5 & $<$1.7 & $<$0.2  \\
   2002/05/22 & {\Chandra}&SGR A$^*$& $<0.23$& $<0.027$ &---&---\\
   2002/05/24 & {\Chandra}&SGR A$^*$& $<0.20$& $<0.023$  &---&---\\
   2002/05/27 & {\Chandra}&SGR A$^*$& 0.13$\pm$0.024 & 0.015$\pm$0.003 &---&---\\
   2002/05/30 & {\Chandra}&SGR A$^*$& $<0.17$& $<0.019$ &---&---\\
   2002/06/03 & {\Chandra}&SGR A$^*$& $<0.21$& $<0.025$ &---&---\\
   \hline
   2001/07/18--19 & 
             {\Chandra} & GCS 16+17 & --- & ---& $<$0.2 & $<$0.03  \\
%            {\Chandra} & GCS 16+17 & $<$0.19& $<$0.027& --- & --- \\
   2002/05/22--06/04 &
             {\Chandra} & SGR A$^*$ & 0.10$\pm$0.014 & 0.012$\pm$0.0016&---&---\\
\hline
\end{tabular}
\medskip

%When the source is not significantly detected, the 3$\sigma$ upper-limit 
%is given.  The uncertainties given are in 1$\sigma$ confidence.

\end{minipage}
\end{table*}

\begin{table*}
 \centering
 %\begin{minipage}{140mm}
 \begin{minipage}{160mm}
  \caption[]{The best-fitting spectral parameters plus 90\% confidence
intervals for the two sources. \label{tbl:fit}
}
  \begin{tabular}{lccccccc}
   \hline
     & Obs. & {\NH}\footnote{Hydrogen column density ($10^{22}${\NHUNIT}).} &
   $\Gamma$ &  ${F_{\rm 6.7}}$\footnote{6.7 keV line flux ($10^{-4}$ ph 
   s$^{-1}$ cm$^{-2}$)} & $E_{\rm 6.7}$\footnote{Line-centre energy 
(keV)}&
   {\FX}\footnote{Observed 2--10 keV flux ($10^{-12}${\FLUXUNIT}).} &
   $\chi^2$/dof \\
   \hline
{\XMMtr} & GC6  & 5.86 (4.95--7.12) & 0.98 (0.73--1.31) &  0.12 (0.03--0.22) 
& 6.65 (6.46--6.74) & 4.5 & 58.9/75 \\
    & SGR A$^*$ & 5.86 (fixed)      & 1.92 (1.07--2.81) & ---               & ---             & 0.010 & 14.9/11 \\
   \hline
{\XMMfe} & GC10 & 12.4 (10.7--14.2) & 2.00 (1.73--2.13) &  1.14  (0.91--1.35) &
6.68 (6.66--6.70) & 3.7 & 44.8/41 \\
   \hline
  \end{tabular}
\medskip

%Numbers in parentheses are 90\% confidence uncertainties for one
%interesting parameter.

\end{minipage}
\end{table*}

\begin{figure*}
\centering
\begin{minipage}{0.67\textwidth}
\centering\hbox{\includegraphics[width=8 cm,angle=270]{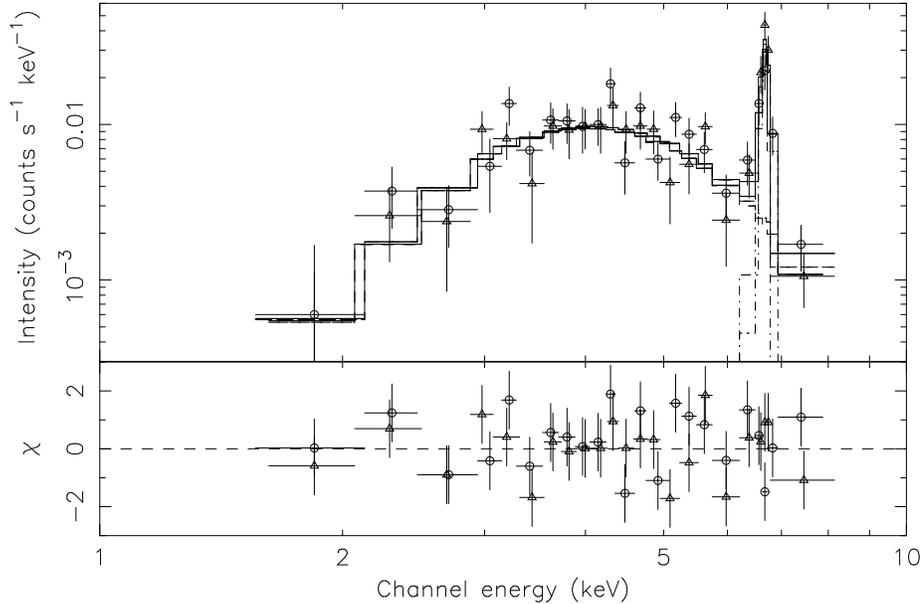}}
\caption{{\XMM} EPIC spectra of {\XMMfe} from the September 2000 (GC10) 
observation. The open circles and triangles represent MOS1 and
 MOS2 data, respectively.  The best-fitting model for each dataset is 
shown as a histogram.
}
\label{fig:xmmfe-spec}
\end{minipage}~~
\end{figure*}

\subsection[]{{\XMMfe}}
\label{sec:XMMfe}

{\XMMfe} is in the field of view for 5 of the
observations listed in Table~\ref{tbl:obslog}. Table~\ref{tbl:flux} 
summarises, in chronological order, the X-ray flux measurements at 
the source position.  The source was detected only in the first 
observation (September 2000) and was at least two orders of magnitude 
fainter some 10 months later.

Within the {\XMMN} GC10 observation the emission from the source was
found to be rather stable.  For example fitting the light curve with a
constant emission model (after background subtraction) for a binning of
32~s and 64~s gives acceptable $\chi^2$ values.  A search for pulsations
using a FFT algorithm provided no evidence for a significant periodic
signal.

We extracted MOS spectra from the GC10 observation
using a $52''$ radius source cell centred on {\XMMfe}.  Corresponding
background spectra were taken from a concentric annulus with inner
and outer radii of $52''$ and $80''$, respectively.  The source was outside
of the field of view of the pn camera in this observation.

Fig.~\ref{fig:xmmfe-spec} shows the EPIC MOS1 and MOS2 spectra after
background subtraction.  A line feature is very prominent between 6--7 keV.  
We fitted the two MOS spectra simultaneously with a model 
comprising a power-law continuum plus a narrow Gaussian line. This 
simple model provides a good match to the observed spectrum with
a $\chi^{2}$(dof) of 44.8 (41). The best-fitting
parameter values are summarised in Table~\ref{tbl:fit} and 
Fig.~\ref{fig:xmmfe-spec} shows 
the corresponding best-fitting model and the fitting residuals.  
The measured line-centre energy, 6.68 (6.66--6.70) keV, is consistent with the dominant
transition being the 6.7-keV K${\alpha}$ line from helium-like iron.
The line equivalent width is remarkably high, namely 2.4 (1.9--2.8)
keV.
The 2--10~keV luminosity is calculated to be 5$\times 10^{34}${\LUMIUNIT},
assuming a distance of 8.0~kpc.   

We also tried a thin-thermal plasma model, allowing the iron abundance
to vary but fixing all the rest of the abundances at the solar values
(Anders \& Grevesse 1989).  The fitting result is found to be acceptable
($\chi^2$/dof=47.5/42).  The best-fitting temperature and iron
abundance are 3.9 (2.6--5.3) keV and 3.1 (1.9--5.1) solar, respectively.

\section[]{Discussion}

Both transient sources have X-ray spectra characterised by
substantial soft X-ray absorption, corresponding to $N_H = 6 
\times 10^{22}${\NHUNIT} in the case of {\XMMtr} and twice this value for 
{\XMMfe}. Given the measured range of column densities associated
with {\GC} features and sources  (Sakano {\etal} 1999; Sakano 2000),
it is perfectly plausible that both sources are located in
the vicinity of the {\GC}.
Thus, the assumption of a distance of 8.0~kpc (Reid 1993) from which 
we infer peak observed X-ray luminosities of 5$\times 10^{34}${\LUMIUNIT} 
would seem to be a reasonable one.  For both sources the lowest observed 
``quiescent'' luminosity is at least two orders of magnitude below
the peak level.

\subsection{Classification of {\XMMtr}}
\label{sec:dis-XMMtr}

{\XMMtr} has a very flat X-ray spectrum with a relatively weak He-like 
iron line (equivalent width of $\sim 200$~eV). This spectral 
description matches two source classes: cataclysmic variables (CVs) in which the
magnetic field of the white dwarf star controls the accretion flow
({\ie}, polars and intermediate polars, see Ezuka \& Ishida 1999)
or high-mass X-ray binaries (HMXBs) in which a neutron star accretes from the wind of a young
massive companion.  The inferred X-ray luminosity in the high state 
of $5\times 10^{34}${\LUMIUNIT} presents a problem for the former
interpretation since this is a factor $\sim$10 times higher than is 
observed in even the brightest magnetic CVs (Verbunt {\etal} 1997;
Ezuka \& Ishida 1999; Pooley {\etal} 2002;  Baskill, Wheatley \& Osborne 2004).  HMXBs are more typically seen
in outburst with luminosities of $L_{\rm X} > 10^{36}$ {\LUMIUNIT} but this
apparent high luminosity preference could well be a selection effect 
(Pfahl, Rappaport \& Podsiadlowski 2002). Similarly the quiescent luminosity of 
1$\times 10^{32}${\LUMIUNIT} is lower than the observed lower limit 
for most HMXBs ($> 5\times 10^{32}${\LUMIUNIT}; Muno {\etal} 2003a).
 Modulation of the X-ray 
light curve arising from the spin period of a wind-accreting neutron 
star can be of relatively large amplitude 
(5\%--50\%) (see Muno {\etal} 2003b).  However the search for an
underlying coherent periodicity in {\XMMtr} is limited by both the
statistics and exposure even when the source was in its high state.
If the equivocal period of 5.2521~s could be shown to be a genuine periodicity, 
then this would lend support to the HMXB scenario.
Also, if the weak diffuse emission seen at and near the location
of  {\XMMtr} could be associated with a star forming region,  
then the HMXB scenario would gain further credence.

\subsection{Classification of {\XMMfe}}

The most interesting spectral property of this source is its extremely 
strong iron line (equivalent width of $\sim$2.4 keV).  
In general, low-mass X-ray binaries (LMXBs) show little evidence for 
line emission, whereas HMXBs sometimes show 
significant Fe lines, presumably emitted in the region of the stellar wind of
the companion star. However an equivalent width of over 1 keV 
is not normally seen in such circumstances.  In addition,
the quiescent X-ray luminosity of {\XMMfe} is unusually low for a HMXB
({\eg} Muno {\etal} 2003).  Active stellar coronal sources such as RS-CVn 
systems, Wolf-Rayet (WR) stars and Young Stellar Objects (YSO) usually have 
much lower luminosities ($\le 10^{32.5}${\LUMIUNIT} at peak) and hence are 
not likely candidates, although they often show an iron line and significant
flares.  Massive WR-OB star binaries may have 2--10 keV X-ray luminosities up to
$10^{35}${\LUMIUNIT}, presumably due to wind-wind collisions (Portegies
Zwart, Pooley \& Lewin 2002).  In addition they often show a strong 
iron K-line with an equivalent width of $\sim$1 keV.  However, here the 
problem is the factor 100 change in luminosity -- for example, an extensive
study covering the different orbital phases shows the variability in the
hard X-ray component in such systems is at most a factor of three 
(Maeda {\etal} 1999).  A background active galaxy, seen through the Galaxy, 
might conceivably be a candidate but again it is very difficult to match
the observations with any known class of object. For example, in Seyfert I
nuclei the measured 6.7-keV line equivalent width is invariably less than
a few hundred eV ({\eg}, Nandra \& Pounds 1994; Reeves \& Turner 2000),
whereas the extreme variability excludes an obscured Seyfert II nucleus 
({\eg} Koyama {\etal} 1989) as a potential counterpart.

The nature of {\XMMfe} is remarkably similar to AX J1842.8$-$0423, which
was discovered in an  {\ASCA} survey of the Scutum arm region in 1996 (Terada
{\etal} 1999).  Terada {\etal} (1999) found that AX J1842.8$-$0423 has a strong 
iron line at 6.8~keV with an equivalent width of $\sim$4~keV.  The spectrum is
well approximated with a thin-thermal plasma model with a temperature of
kT$\sim$4~keV and heavy-metal abundance of $\sim$3 solar.  AX
J1842.8$-$0423 also showed transient behaviour on a time-scale of
less than half a year. In addition there are two other sources which 
have an extremely strong iron line: AX J2315$-$592 (CP Tuc; Misaki {\etal} 
1996) and RX J1802.1+1804 (V884 Her; Ishida {\etal} 1998).  Both of
the latter sources have been identified as polars ({\ie} AM Her type
CVs).

Terada {\etal} (1999, 2001) concluded that the three sources above are 
probably close binaries involving a magnetised white dwarf viewed from 
a pole-on inclination.  In fact, although many CVs show an
iron line, the equivalent width of such features is typically 
a couple of hundred eV or less, or the iron abundance 
is generally sub-solar ({\eg} Ezuka \& Ishida 1999).
However, Terada {\etal} (1999, 2001) argued that
the observed strong iron line is interpreted as due to 
line-photon collimation in the accretion column of the white dwarf, as a
result of resonance scattering of line photons, which may occur in the
case of a pole-on inclination.

\begin{table*}
 \centering
% \begin{minipage}{140mm}
 \begin{minipage}{160mm}
  \caption[]{The spectral parameters for four sources with strong iron 6.7-keV 
emission.
  \label{tbl:polar}
  }
  \begin{tabular}{lccccccccc}
   \hline
Name            & {\NH}               & $kT$  &
   $E_{\rm 6.7}$\footnote{Centre energy ($E_{\rm 6.7}$) and equivalent width (EW$_{\rm 6.7}$) of the 6.7-keV line.} &
                            EW${_{\rm 6.7}}^{a}$ &
                              $Z_{\rm Fe}$\footnote{Iron abundance relative to the solar abundance ratio.} &
                                {\LX}\footnote{The 2--10 keV luminosity
   (in the high state if the source shows the transient activity).  $d$ 
indicates the distance; {\eg}, $d_8$ is the distance in units of 8~kpc.} &
                                  ID\footnote{Are there any optical/IR identifications?} &
                                    Tr?\footnote{Has X-ray transient activity ever been observed?} &
                                      Ref.\\ 
                & ($10^{22}${\NHUNIT})& (keV) &
   (keV)                  & (keV)               &                     & ({\LUMIUNIT})
    & &\\
   \hline
AX J2315$-$592  & $<$0.07             &16.5$^{+3.0}_{-3.1}$ &
   6.84$^{+0.13}_{-0.09}$ & 0.9$^{+0.3}_{-0.2}$ & 1.93$^{+0.56}_{-0.43}$&
   $1\times 10^{32} {d_{0.2}}^2$ & Yes & No  & 1,2,3 \\ %FX=27e-12
RX J1802.1+1804\footnote{Results for a
   single absorption and single temperature model.}
                & $\sim$0.01          & $\sim$0.9           &
   6.55$^{+0.09}_{-0.08}$ & 4.0$^{+1.7}_{-1.7}$ & 1.3 -- 12           &
   $6\times 10^{29} {d_{0.1}}^2$ & Yes & No  & 4,5 \\ %FX=0.5e-12
AX J1842.8$-$0423 & 3.9$^{+1.1}_{-1.1}$ & 5.1$^{+5.0}_{-1.9}$ &
   6.78$^{+0.10}_{-0.13}$ & 4.0$^{+1.0}_{-0.5}$ & 3.0$^{+4.3}_{-0.9}$ &
   $1.6\times 10^{34} {d_{5}}^2$ & No  & Yes & 6 \\ %FX=5.2e-12
{\XMMfe}        &12.4$^{+1.8}_{-1.7}$ & 3.9$^{+1.4}_{-1.3}$ &
   6.68$^{+0.02}_{-0.02}$ & 2.4$^{+0.4}_{-0.5}$ & 3.1$^{+2.0}_{-1.2}$ &
   $5\times 10^{34} {d_{8}}^2$   & No  & Yes & 7 \\ %FX=3.7e-12
   \hline
\end{tabular}
\medskip

References -- 1: Misaki {\etal} 1996; 2: Thomas \& Reinsch 1996; 3:
  Ramsay {\etal} 1999; 4: Ishida {\etal} 1998; 5: Greiner {\etal} 1998;
  6: Terada {\etal} 1999; 7: this work.
\end{minipage}
\end{table*}

Table~\ref{tbl:polar} summarises the properties of the three
sources considered by Terada {\etal} in comparison to those of
{\XMMfe}.
The extremely strong iron line may be explained with the model of Terada
{\etal}  However its high luminosity of $\sim 10^{34}${\LUMIUNIT}
still remains a major problem as discussed in the previous section (\S\ref{sec:dis-XMMtr}).

Given the uncertainty pertaining to the CV scenario, what other possibilities
are there? It is plausible that this source represents a new type of 
neutron star binary which, for some reason, is relatively efficient at producing 
strong iron lines.  In this case, the transient nature and high luminosity would 
be easily explained.  Alternatively, a massive WR-OB star binary in an extremely 
eccentric orbit might conceivably give rise to transient-like activity.
Clearly, it remains to be demonstrated whether {\XMMfe} corresponds to
a known class of source in an extreme configuration or to an entirely
new class of object.

\subsection{The origin of the {\GC} 6.7-keV line emission}

Recent deep observations with {\Chandra} have shown that the 
bulk of the hard X-ray emission associated with the inner Galactic Plane
is not due to the summed emission of faint Galactic sources, at least 
down to a limiting source luminosity of $L_X \approx 10^{31}$ {\LUMIUNIT} 
(Ebisawa {\etal}  2001). The demonstration that this feature, known
as the Galactic Ridge ({\eg} Worrall {\etal} 1982; Warwick {\etal} 1985;
Koyama {\etal} 1986 ; Yamauchi \& Koyama 1993; Kaneda {\etal} 1997; 
Valinia \& Marshall 1998), is truly diffuse in origin represents a major step 
forward. However, our understanding of the processes that produce both the 
hard continuum and the associated line emission from
highly ionized ions, most notably helium-like iron at 6.7-keV, 
remains incomplete ({\eg} Tanaka {\etal} 2000; Tanaka 2002; Valinia {\etal} 2000).
Interestingly the hard X-ray emission observed from the
{\GCR} has very similar spectral characteristics to that
of the  Galactic Ridge (Koyama {\etal} 1996; Kaneda {\etal} 1997; 
Tanaka {\etal} 2000; Tanaka 2002) except for its much higher surface
brightness in the {\GCR}.

The high concentration of point sources observed within $9'$ of the {\GC}
by {\Chandra} (Muno {\etal} 2003a) suggests the possibility that, at least in 
this region, point sources may contribute significantly to the total observed 
emission. Particularly intriguing in this respect is the fact that the 
integrated spectrum of faint resolved sources is extremely hard 
($\Gamma \approx 0.8$) with a strong He-like iron line 
(equivalent width of $\sim 400$ eV) (Muno {\etal} 2004b). However, 
investigation of the  {\Chandra} ${\log} N$-${\log} S$ curve for resolved
X-ray sources shows that such sources account for only $\sim$10\% of 
the total 2--8 keV emission (Muno {\etal} 2004a). If unresolved sources make up 
the deficit, the requirement is for a total of $2 \times 10^{5}$ sources
within 20 pc ($9'$) of Sgr A* with a typical luminosity
$L_{\rm X}\sim 10^{30}$ {\LUMIUNIT}. In fact the implied 
source number density  substantially exceeds the estimated $10{^4}$ CVs 
in the region and, together with the additional luminosity and spectral 
constraints, matches no known population of sources (Muno {\etal} 2004a).

Our observation of {\XMMfe} suggests one further possibility. This
source has very strong 6.7-keV iron line emission and in fact the line 
equivalent width of $\sim 2.4$ keV is roughly 6 times that of the
average faint source resolved by {\Chandra} to a limiting luminosity of
$\sim 10^{31}${\LUMIUNIT}. Might therefore similar sources explain all 
or part of the 6.7-keV line emission without necessarily also accounting
for the bulk of the hard continuum?

We estimate from unpublished {\XMMN} data that the total 6.7-keV iron-line 
flux from the region within $9'$ of the {\GC} is $\sim 7 \times 10^{-4}$ 
photon~s$^{-1}$~cm$^{-2}$, excluding the contribution of the Sgr A East 
SNR (Maeda {\etal} 2002; Sakano {\etal} 2004a,~b). The 6.7-keV line flux of 
{\XMMfe} in its high state is 1.1$\times 10^{-4}$ photon~s$^{-1}$~cm$^{-2}$ 
(Table~\ref{tbl:fit}). However, no other similar source (in terms 
of the line properties) is seen at the same flux level in the {\XMMN} {\GC} 
Survey. Also since we have established that the source is a transient,
it is clear the presence of relatively large numbers of such sources
in quiescence is of principle concern.
If we assume that these transient sources keep their spectral shape, 
including the large equivalent width of their 6.7-keV line, in their 
quiescent state, then the number of {\XMMfe}-type sources required 
to explain the whole 6.7-keV line intensity of the central region
is $\sim 30000~{L_{31}}^{-1}$, where ${L_{31}}$ is the typical quiescent
luminosity in units of 10$^{31}${\LUMIUNIT}. As expected the integrated
emission of such sources would account for only $\sim 15\%$ of the 2--8 keV 
continuum and might be a poor match to the slope of the observed spectrum 
(this is uncertain since {\XMMfe} has a rather soft continuum form but also 
a substantial column density). 

Unfortunately this scenario looks barely viable. If $L_{31} > 10 $
then such sources should be appearing in large numbers (with prominent 
6.7 keV lines) in the {\Chandra} surveys.  At $L_{31} \approx 1$ 
such sources should again be appearing as faint resolved 
{\Chandra} sources, yet the measured 6.7-keV line equivalent width
of the integrated resolved-source spectrum limits their contribution 
to $\sim 15\%$.  Conversely, if the quiescent state of such transients 
is at fainter levels (\ie $L_{31} < 0.3$), then the required
source density becomes excessive, particularly if there is an association 
with CVs.

\section{Conclusion}

We have discovered two X-ray transient sources, {\XMMtr} and {\XMMfe}, in the
{\GCR} with {\XMMN}.  The observations of {\XMMN} and {\Chandra}
revealed that the flux of the two sources changed by more than two orders of
magnitude in less than a year.  Their heavy absorption in the high state
gives a strong indication that both the sources are located in the {\GCR}.  
The peak
fluxes of both the sources are $\sim 4\times 10^{-12}${\FLUXUNIT}, equivalent
to an X-ray luminosity of $\sim 5\times 10^{34}${\LUMIUNIT}, suggesting a new
transient population with a peak luminosity up to three orders
of magnitude below that typical of X-ray transients in our Galaxy. 
{\XMMfe} shows a strong emission line at 6.7~keV, presumably a
K-line from helium-like iron, with an equivalent width of $\sim$2.4 keV.
The nature is very similar to AX~J1842.8$-$0423.  However,
no known class of sources can well explain all their characteristics.
{\XMMtr} has a very flat spectrum
($\Gamma\sim$1.0) with a possible weak line from He-like iron.  It is
the most likely a neutron star or black hole binary, possibly with a
high-mass companion.

\section*{Acknowledgments}

The authors would like to express their thanks to all those who have
contributed to the successful development and operation of {\XMMN}. We
are grateful to J. Tedds, D. Baskill and the anonymous referee for their valuable comments.  In
addition we should like to acknowledge the help of many colleagues at
Leicester, especially R. Saxton, S. Sembay, I. Stewart and
M.J.L. Turner, on matters relating to the EPIC calibration and the use
of the {\sc SAS}.

%\appendix

%\bsp

\label{lastpage}

\end{document}